\documentclass[manuscript,nonacm]{acmart}
\usepackage{subfigure}
\usepackage{paralist}
\usepackage{enumitem}
\usepackage{mdframed}
\usepackage{csquotes}
\AtBeginDocument{%
  }

\begin{document}

\title{Explainable AI in Usable Privacy and Security: Challenges and Opportunities}

\author{Vincent Freiberger}
\affiliation{%
  \institution{Center for Scalable Data Analytics and Artificial Intelligence (ScaDS.AI) Dresden/Leipzig}
  \city{Leipzig}
  \country{Germany}
}
\affiliation{%
  \institution{Leipzig University}
  \city{Leipzig}
  \country{Germany}
}
\email{freiberger@cs.uni-leipzig.de}

\author{Arthur Fleig}
\affiliation{%
  \institution{Center for Scalable Data Analytics and Artificial Intelligence (ScaDS.AI) Dresden/Leipzig}
  \city{Leipzig}
  \country{Germany}
}
\affiliation{%
  \institution{Leipzig University}
  \city{Leipzig}
  \country{Germany}
}
\email{arthur.fleig@uni-leipzig.de}

\author{Erik Buchmann}
\affiliation{%
  \institution{Center for Scalable Data Analytics and Artificial Intelligence (ScaDS.AI) Dresden/Leipzig}
  \city{Leipzig}
  \country{Germany}
}
\affiliation{%
  \institution{Leipzig University}
  \city{Leipzig}
  \country{Germany}
}
\email{buchmann@informatik.uni-leipzig.de}

\renewcommand{\shortauthors}{Freiberger et al.}

\def\AF{\textcolor{blue}}

\begin{abstract}

Large Language Models (LLMs) are increasingly being used for automated evaluations and explaining them. However, concerns about explanation quality, consistency, and hallucinations remain open research challenges, particularly in high-stakes contexts like privacy and security, where user trust and decision-making are at stake. In this paper, we investigate these issues in the context of PRISMe, an interactive privacy policy assessment tool that leverages LLMs to evaluate and explain website privacy policies. Based on a prior user study with 22 participants, we identify key concerns regarding LLM judgment transparency, consistency, and faithfulness, as well as variations in user preferences for explanation detail and engagement. We discuss potential strategies to mitigate these concerns, including structured evaluation criteria, uncertainty estimation, and retrieval-augmented generation (RAG). We identify a need for adaptive explanation strategies tailored to different user profiles for LLM-as-a-judge. Our goal is to showcase the application area of usable privacy and security to be promising for Human-Centered Explainable AI (HCXAI) to make an impact.

\end{abstract}

\keywords{Usable Privacy, LLMs, Explainability, Presented at the Human-centered Explainable AI Workshop (HCXAI) @ CHI 2025, DOI: \href{https://doi.org/10.5281/zenodo.15170451}{10.5281/zenodo.15170451}}


\maketitle

\section{Introduction}
\label{sec:intro}
With increasing cybersecurity threats -- e.g., through the growing number of devices introducing more vulnerabilities -- and privacy threats -- such as user data being used to train AI models -- empowering and informing users about these risks has become increasingly important. This is underscored by the fact that most cybersecurity incidents are facilitated by human error~\cite{walker2020threats}.
In the field of usable privacy and security, model judgements and explanations and the effective handling of hallucination scenarios are particularly important, as users' sensitive data is at stake and misguided judgments can have severe consequences. A promising approach, which has already gained traction in other research fields and holds potential for usable privacy and security, is utilizing LLM-as-a-judge~\cite{zheng2023judging,gu2024survey}. This approach involves instructing an LLM to evaluate a given document and assign scores. However, despite its potential, key challenges remain unresolved. The explainability of LLM-generated scores, beyond instructing the LLM to self-report its reasoning, remains underexplored. Additionally, tailoring these explanations to different user profiles to enhance their effectiveness in the context of usable security and privacy has yet to be investigated. Furthermore, hallucinations in LLM-as-a-judge scenarios remain largely unexamined, and mitigation strategies have yet to be explored.

Our specific use case motivating this research is PRISMe, an interactive privacy policy assessment tool we designed~\cite{freiberger2025prisme}. It employs an LLM-as-a-judge approach to evaluate and rate privacy policies, providing users with an overview while incorporating a conversational component that allows them to ask questions about the policy. Understanding how users interact with explanations provided by such a tool is crucial to ensure understanding, trust, and good decision making~\cite{kim2024human}, as well as determining which types of explanations are most helpful for different user groups. Our goal is to maximize users' privacy and security awareness without exposing them to unintended risks, such as misinterpretation or a false sense of security.

While multiple benchmarks assess LLM hallucinations in question-answering settings~\cite{huang2025survey} and allow to transfer insights to PRISMe, such an understanding is still lacking for LLM-as-a-judge as in our initial policy assessment. While our previous work~\cite{freiberger2025prisme} has focused on discussing issues and potential solutions for the interactive chat component, we aim to focus on the LLM-as-a-judge component here.

In particular, we identify the following areas for future research and outline our initial thoughts on addressing them:
\begin{compactitem}
    \item Investigating how different user types interact with LLM-generated explanations in the context of privacy and security (addressing individual factors influencing explainability needs of different users building on Li et al.~\cite{li2024exploring}).
\item Customizing LLM-as-a-judge assessments for different user profiles to enhance interpretability and effectiveness.
\item Continuing the trajectory laid out by Abdul et al.~\cite{abdul2018trends} and work from previous Human-Centered Explainable AI (HCXAI) research like~\cite{kim2024human,ehsan2020human,datta2023s} to create user-tailored, interactive explanations that in our use case help minimize exposure to security and privacy threats.
\item Understanding and detecting hallucinations in an LLM-as-a-judge setting and developing mitigation strategies tailored to privacy and security contexts (addressing the when to trust). This continues the research theme of appropriate trust calibration in the HCXAI community~\cite{graichen2022facilitate,hemmer2022effect,liao2021human}.
\end{compactitem}

Our goal is to highlight the promising application of usable security and privacy to the HCXAI-community, where studying model explanations and LLM hallucinations from a human-centered perspective involving personalization can have a significant positive impact on lay users.
\section{Related Work}\label{sec:related}
\textbf{Challenges with Privacy Policies}
\label{sec:policy}
\\
Privacy policies aim to reduce the \emph{information asymmetry} between service providers and users~\cite{malgieri2020concept,Zaeem2020}. However, they are often designed for legal compliance, with dense, complex, and lawyer-centric language~\cite{schaub2017designing}. 
This makes it difficult for users to make informed decisions about their privacy online~\cite{windl2022automating,mhaidli2023researchers}. Legal regulations like the GDPR~\cite{eu2016regulation} do not prevent persuasive language, which makes unethical practices even harder to detect~\cite{pollach2005typology,belcheva2023understanding}. Additionally, such regulations may create a false sense of trust and security. 
Users rarely read and typically don't understand privacy policies~\cite{reidenberg2015disagreeable,steinfeld2016agree}, which leads to informational unfairness~\cite{freiberger2024legal}.
Generative AI~\cite{lee2024deepfakes} and Augmented Reality complicate data management practices further, thereby exacerbating transparency issues~\cite{becher2021law,belcheva2023understanding,transparency}.
\\
\textbf{LLM-based privacy policy assessment}
\label{sec:assistants}
\\
LLM-based privacy policy assessment has shown to be as effective as NLP methods in extracting key aspects from privacy policies, such as contact information and third parties~\cite{rodriguez2024large}.
ChatGPT-4~\cite{OpenAI2024} offers performance and adaptability in answering privacy-related questions~\cite{hamid2023genaipabench}, which motivated our use of LLMs in a interactive privacy policy assessment tool. 
Privacify~\cite{woodring2024enhancing} is a browser extension performing information extraction and abstract summarization of policies addressing compliance and data collection information. While it lacks interpretation, customization, and interactive features, it further motivated us to utilize the LLM-as-a-judge paradigm in a tool that interprets and explains privacy policies and allows users to interact with the provided information~\cite{freiberger2025prisme}. We explain our tool in Section~\ref{sec:sys}.
\\
\textbf{LLM-as-a-judge}
\\
The LLM-as-a-judge paradigm originates from the practice of using LLMs to evaluate other LLM-generated outputs~\cite{zheng2023judging}. This approach has been widely applied in text evaluation tasks, where LLMs rate and score input text~\cite{gu2024survey,li2024generation} even evaluated in a RAG setting~\cite{xu2025does}.
Despite becoming a new standard in NLP evaluation, LLM-as-a-judge is subject to biases~\cite{chen2024humans,ye2024justice,xu2025does}, its judgments do not always align with human assessments~\cite{bavaresco2024llms,szymanski2025limitations}, and the presented reasoning sometimes is flawed~\cite{xu2025does}, motivating research on LLM explanations.
\\
\textbf{LLM Explanations}
\\
In the field of HCXAI, LLMs are increasingly used to make traditional Explainable AI (XAI) methods' outputs more understandable for humans~\cite{zytek2024llms}. LLMs' explanations of their own outputs cannot be seen as mechanistically reflecting their inner workings but rather as samples of post-hoc rationalizations~\cite{sarkar2024large}. However, step-by-step self-explanations of the model to derive an output can significantly improve output quality~\cite{kojima2022large,liu2019towards}. 
To ground explanations, citation-based explanations could link to evidence in the context~\cite{sarkar2024large}. LLMs explaining themselves could help users to critically reflect, justify, and evaluate if prompted accordingly~\cite{sarkar2024ai}.
\\
\textbf{Hallucinations}
\\
Hallucinations can be defined as the generation of nonsensical outputs or output that is unfaithful to the provided source content~\cite{maynez-etal-2020-faithfulness}.
Huang et al.\ divide hallucinations into ones concerning faithfulness, i.e., whether outputs are faithful to instructions and context given, and ones concerning factuality, i.e., whether outputs are factually correct~\cite{huang2025survey}. 

\section{Use Case: Privacy Risk Information Scanner for Me (PRISMe)}
\label{sec:case}
We investigate LLM explanations and hallucinations in the application area of usable privacy and security.
To this end, we utilize the exemplary use case of PRISMe, an interactive privacy policy assessment tool. This section explains PRISMe and a previous user study we conducted with it~\cite{freiberger2025prisme}.
\subsection{System}
\label{sec:sys}
\begin{figure*}[htb]
    \centering
    \includegraphics[width=0.9\textwidth]{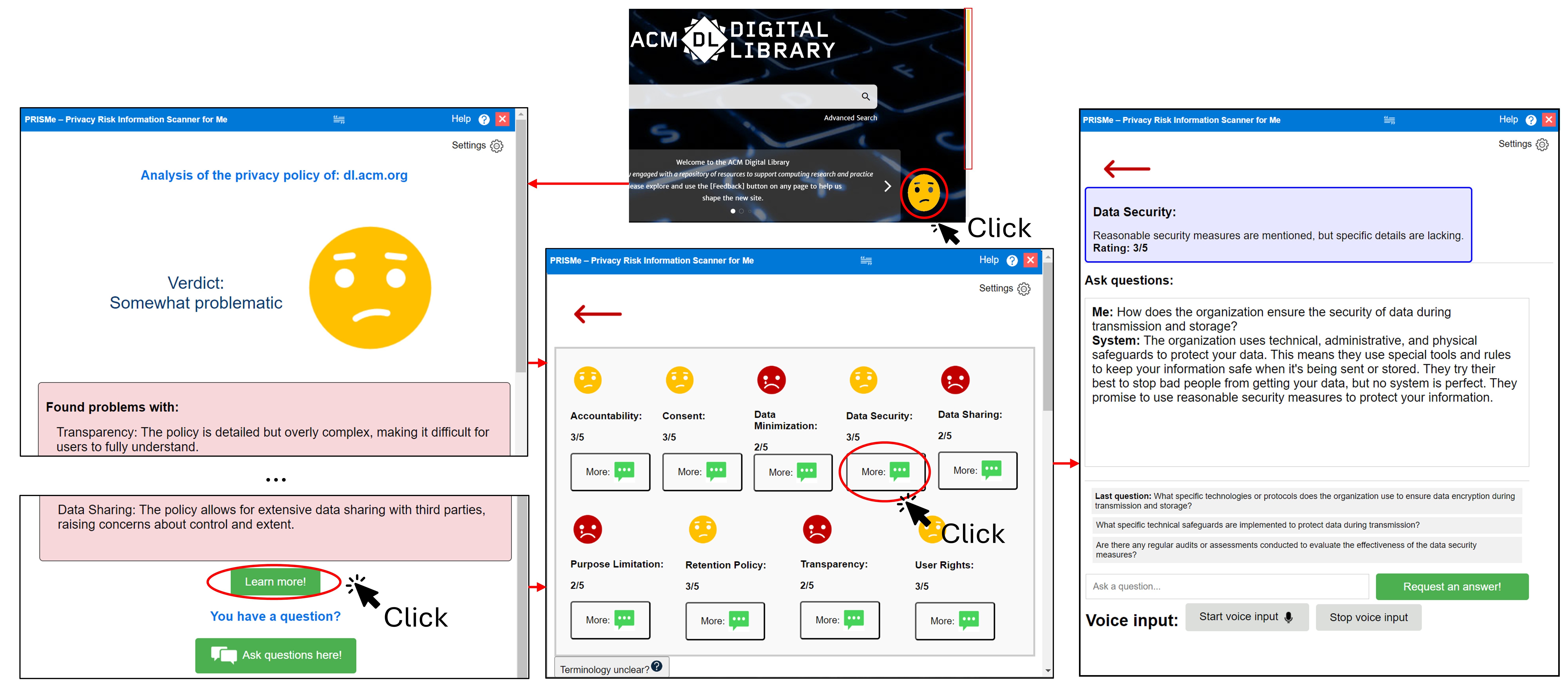}
    \caption{When the user visits a website, our prototype evaluates the privacy policy in the background and displays privacy alerts via colored scrollbars and a point-of-entry smiley icon (top middle). Clicking the smiley opens an Overview Panel (left) summarizing key privacy issues, with navigation to a Dynamic Dashboard and chat interface. The dashboard (bottom middle) provides detailed policy evaluation criteria, which users can chat about (right) by clicking the respective "More" button.}
    \Description{The image provides a multi-step walkthrough of using Privacy Risk Information Scanner for Me on the ACM Digital Library (dl.acm.org) website to review its privacy policy. The sequence of interactions is illustrated with arrows showing clicks and actions. Here's a description of each section:
    Top Middle
    The ACM Digital Library homepage is displayed with the logo, search bar, and a concerned face icon (indicating a "Somewhat problematic" privacy policy verdict).
    There is an arrow next to the smiley, showing to click on it for more information on the policy.
    Left
    This shows the PRISMe - Privacy Risk Information Scanner displaying the privacy policy's analysis for dl.acm.org, with a verdict of "Somewhat problematic." The highlighted issues are transparency and data sharing.
    An arrow points to a clickable buttons labeled "Learn more!" above another button "Ask questions here!" to get further details.
    Bottom Middle
    After clicking "Learn more," a more detailed dashboard breakdown of the policy's components is shown, including various categories such as Accountability, Consent, Data Minimization, Data Security, Data Sharing, Purpose Limitation, and others.
    Each category is rated with green, yellow or red colored smiley faces, showing satisfaction levels (e.g., Data Security is rated 2/5 with a red sad smiley). An arrow points to the "More" button under the Data Security section.
    Right
    After clicking "More" on the Data Security section, details on the site's security measures are displayed. It mentions that the policy provides "reasonable security measures" but lacks specifics, giving it a 3/5 rating.
    Below, there is a Q and A interface where the user asks, "How does the organization ensure the security of data during transmission and storage?" The system responds, explaining the use of technical, administrative, and physical safeguards to protect data, though acknowledging no system is perfect.
    The bottom of the section has a text input field for additional questions.
    In summary, the image illustrates how a user can interact with the PRISMe scanner to assess the privacy policy of the ACM Digital Library, focusing on data security and receiving additional insights through interactive features.}
    \label{fig:ui}
\end{figure*}

Figure~\ref{fig:ui} illustrates PRISMe, the tool we developed. When users visit any website, PRISMe fetches its privacy policy with our scraper utilizing headless browser automation. We automatically analyze its privacy policy, highlighting concerns using smiley icons (Figure~\ref{fig:ui}, top middle). Following the LLM-as-a-judge paradigm, the tool dynamically selects and evaluates criteria on a 5-point Likert scale for policy assessment (see prompts in Appendix~\ref{sec:appendix}) utilizing GPT-4o~\cite{OpenAI2024}.
The smiley icon (green, yellow, red) provides a quick summary of the policy’s overall rating. Clicking it opens the Overview Panel (Figure~\ref{fig:ui}, left), which summarizes key issues, with links to the dashboard and chat interfaces for further exploration.
The Dashboard Panel (Figure~\ref{fig:ui}, bottom middle) displays detailed scores for each assessment criterion, rated with corresponding smiley icons. Below the dashboard, an explanation of each score is provided. Users can further explore specific criteria via the Criteria Chat (Figure~\ref{fig:ui}, right) or use the General Chat for broader privacy-related inquiries.
Settings, accessible via a cogwheel icon, allow users to customize the length (short, medium, long) and complexity (beginner, basic, expert) of chat responses and policy assessments, tailoring the tool to their preferences and technical expertise.

\subsection{User Study}
Utilizing the prototype described above, our prior work contains a user study with 22 participants to evaluate it~\cite{freiberger2025prisme}. Participants used the prototype in three scenarios. In the first scenario, they were instructed to take their time exploring the privacy policies of Paypal and focus.de (a popular German news media website) using PRISMe. The aim was to investigate understanding and awareness. The second scenario addressed participants' efficiency using PRISMe: participants compared four online bookshops (Amazon, Hugendubel, Buchkatalog and Kopp), considering the privacy policy evaluations provided by our prototype. The last scenario allowed participants to use our tool on any website of their choice (37 different websites were visited). The aim was to investigate understanding, usability, and awareness.

User studies investigating human perception of XAI often focus on related concepts like trust, understanding, usability, or human-AI collaboration performance~\cite{rong2023towards}.
Comments made by participants during the scenarios and in semi-structured interviews on how they perceived assessment transparency and trustworthiness of the tool motivated us to investigate explainability of LLM-as-a-judge and to connect to the HCXAI community. 
Some participants wanted clear policy evidence to justify specific ratings. They found that the LLM-generated explanations were often overly generic, with little distinction between descriptions for good and bad ratings.
Another issue was inconsistency in criteria selection. Participants noted that the LLM interpreted similarly named criteria differently between policies or selected entirely different criteria for similar policies.
Additionally, some participants observed incoherence between chat responses and LLM-provided ratings—the explanations justifying ratings sometimes did not align with the sentiment expressed in chat responses. Participants still found LLM provided explanations helpful. 

Our study also revealed different user profiles regarding their usage behavior. Besides usage behavior, our profiles take into account participants' self-reported prior knowledge and confidence in their understanding of typical privacy policies and general data protection as collected in a questionnaire before the study. We suspect these groups differ in the way they process LLM explanations and potential hallucinations, raising different requirements for explanation. 
\begin{compactitem}
    \item \textit{Targeted Explorers:} Users with prior knowledge who critically inspect evaluations. They prefer specific, extensive explanations and clearly defined evaluation criteria.
\item \textit{Novice Explorers:} Users with little prior knowledge who rely on tools like PRISMe for guidance. Their learning goals emerge as they explore. They engage actively and take their time to process information.
\item \textit{Information Minimalists:} Users who seek quick, high-level summaries with minimal engagement. They prefer concise overviews rather than interactive exploration.
\end{compactitem}

These diverse information-seeking behaviors require tailored strategies to effectively communicate explanations and foster awareness. This offers ground for fruitful discussions with the HCXAI community.

\section{Discussion}
This section discusses LLM explanation and hallucination issues, balancing transparency, trust and critical thinking, and the trade-off between explainability and privacy. We take into account the user profiles we have identified.
\subsection{Investigating the LLM Explanations and Hallucination Issues}
One approach to increase the transparency in an LLM-as-a-judge approach like ours is to reduce the degrees of freedom the LLM has. A fixed criteria catalog with precise definitions and requirements for assigning individual scores could be created. This would likely also improve the judgment accuracy~\cite{xu2025does}.
A drawback of such fixed criteria definitions is that they may be too rigid and unsuitable to accurately evaluate different risk profiles. This could pose risks to \textit{Information Minimalists}, as they mostly rely on initial ratings that in such a case may miss critical information. Also, providing this context does not ensure the LLM actually follows it upon inference. 
Utilizing output token uncertainty for tokens on the scoring may help judge the LLM's confidence in its judgement~\cite{varshney2023stitch}. Multiple assessments may be sampled to see where variance in assigned scores occurs may help quantify confidence~\cite{manakul2023selfcheckgpt}. Communicating such confidence levels supports all user profiles, particularly \textit{Information Minimalists}.
Visualizing attention may also help as a proxy for explanations of what tokens contributed how much to the output~\cite{10541203}. While \textit{Targeted Explorers} are likely happy about additional model explainability to calibrate their level of trust, not overwhelming other user groups with too much additional information is important. Investigating this trade-off could continue the debate on personalizing explanations.

Hallucination scenarios in LLM-as-a-judge use cases have yet to be investigated. The majority of issues may be faithfulness-related. The LLM may ignore the actual policy at hand for evaluation and refer to generic themes. This is supported by the sometimes rather generic explanations the LLM provides for its evaluation. 
Providing clearly defined criteria definitions in this context may help to translate the evaluation of each individual criterion into a RAG problem. As a benefit, explanations for evaluations are probably more specific and can refer to evidence from the policy, however at large computational costs with multiple RAG-queries being performed for the different criteria. 
Sampling multiple assessments and checking them for internal consistency is also a common hallucination detection strategy~\cite{manakul2023selfcheckgpt,huang2025survey} and could help identify unwanted hallucinated evaluation criteria.
Communicating specifics and citing evidence effectively supports \textit{Targeted Explorers} to build trust in provided explanations.

Regarding factuality, one problem is that the LLM fabricates actually irrelevant evaluation criteria. Factuality of the Likert-scale ratings is rather difficult to judge as it is not entirely clear what a policy rated with a 4 instead of a 3 on the criterion data minimization should do better. Assigning Likert-scale ratings is always prone to subjectivity and there is no ground truth. The often rather generic self-explanations provided by the LLM for its rating might not suffice. Here, we arrive at a similar conclusion to Sarkar~\cite{sarkar2024large} and call for future research, particularly in the LLM-as-a-judge setting.

\subsection{Balancing Transparency, Trust, and Critical Thinking}
LLM-generated explanations should help users make informed decisions by enabling an appropriate level of reliance instead of blindly trusting the system’s judgments~\cite{hemmer2022effect}. While explanations can mitigate informational unfairness~\cite{freiberger2024Fair}, they should also encourage users to reflect critically on privacy policies rather than passively accepting automated evaluations.
Different user profiles require varying levels of customization in explanations~\cite{kim2024human}. Future research should explore how to personalize explanations effectively while mitigating biases in LLM-as-a-judge frameworks.

A key challenge in the LLM-as-a-judge approach is aligning its outputs and explanations with human judgment. While lay users often show high agreement with LLMs in expert domains, experts tend to disagree more frequently~\cite{szymanski2025limitations}. Experts value factual accuracy, up-to-date and evidence-based content, and prefer concise, precise, and unambiguous language. In contrast, LLMs often produce verbose outputs with generalized reasoning and limited expert-level rationale.
Szymanski et al.~\cite{szymanski2025limitations} found that prompting LLMs with expert personas -- like we do -- can partially reduce this misalignment. However, this comes at the cost of reduced alignment with lay users, who generally prefer less technical language. Incorporating data protection experts into the loop for screening sample assessments to provide clearer guidelines to the LLM judge combined with evidence-based evaluations via RAG should help. To avoid user overwhelm, particularly for \textit{Information Minimalists}, the prompt design needs to counter LLMs' tendency to favor long and comprehensive responses in complex language.
Even though rare, LLM-generated explanations also raise concerns about subjectivity and the risk of confidently justifying incorrect answers~\cite{kunz2024properties}. Despite this, they often align well with human explanations.
Their tendency toward illustrative examples and selectivity~\cite{kunz2024properties} can be seen more as a feature than a bug as it helps users to build an understanding without getting overwhelmed, given the information presented by the LLM includes the most important aspects.
The alignment with human explanations is found to likely increase trust, which becomes problematic considering convincing justifications for wrong answers. Particularly \textit{Information Minimalists} and \textit{Novice Explorers} are likely to be misled. To what extent forcing an LLM to verify its answer in a separate step, or to incorporate expert reviews, is an open question, where the HCXAI community could contribute significantly.

\subsection{Balancing Privacy and Explainability}
There is a tension between privacy and the quality of the explanation provided to users. Running smaller, local LLMs offers improved privacy around user interests and preferences, at the cost of reduced output quality in both the judge and conversational components. 
By increasing the transparency and quality of the explanations of the LLM-as-a-judge component, there is an increasing risk of service providers slightly modifying their privacy policies to take advantage of lacking adversarial robustness in the LLM and receive better scores without improving their data protection practices. Lakshminarayanan and Gautam~\cite{lakshminarayanan2024balancing} already noted such adversarial robustness trade offs for Explainable AI in general. 
Personalizing assessments requires users to directly or indirectly disclose aspects of their identity, which can compromise privacy, especially if such data is processed by cloud-based models.
Inspired by Lakshminarayanan and Gautam~\cite{lakshminarayanan2024balancing}, we suggest giving the user control by making automated personalization opt-in to protect privacy.


\section{Conclusion}
\label{sec:conclusion}

As the field of usable privacy and security increasingly relies on opaque LLMs, the need for human-centered explainable AI (HCXAI) approaches to enhance transparency is more critical than ever. Intransparent LLM-generated judgments can undermine user trust and lead to errors, posing significant risks when sensitive user data is at stake. Our work highlights key research gaps to address, including user-adaptive explanations, improved interpretability, and robust strategies for detecting and mitigating hallucinations in LLM-as-a-judge frameworks. Bridging these gaps is essential to ensuring that AI-driven privacy tools are not only effective but also trustworthy and user-centric.

\bibliographystyle{ACM-Reference-Format}
\bibliography{literature}
\appendix
\section{Appendix}
\label{sec:appendix}
\subsection{Prompting Approach for Initial Assessment Generation}

    Your output must be a maximum of 600 words long! You are an expert in data protection and a member of an ethics council. You are given a privacy policy. Your task is to uncover aspects in data protection declarations that are ethically questionable from your perspective. Proceed \textbf{step by step}:

    \begin{enumerate}
        \item \textbf{Criteria:} From your perspective, identify relevant ethical test criteria for this privacy policy as criteria for a later evaluation. When naming the test criteria, stick to standardized terms and concepts that are common in the field of ethics. Keep it short! 
        \item \textbf{Analysis:} Based on this, check for ethical problems or ethically questionable circumstances in the privacy policy. 
        \item \textbf{Evaluation:} Only after you have completed step 2: Rate the privacy policy based on your analysis regarding each of your criteria on a 5-point Likert scale. Explain what this rating means. Explain what the ideal case with 5 points and the worst case with one point would look like. The output in this step should look like this: 
        [Insert rating criterion here]: [insert rating here]/5 [insert line break] 
        [insert justification here]
        
        \item \textbf{Conclusion:} Reflect on your evaluation and check whether it is complete.
    \end{enumerate}
    
    Important: Check for errors in your analysis and correct them if necessary before the evaluation. You must present your approach clearly and concisely and follow the steps mentioned. Your output must not exceed 600 words.

\subsection{Prompting Approach for Chat Answer Generation}
\label{sec:chat_prompt}

\textbf{System prompt criteria chat: }Keep it short! Privacy policy: <Privacy policy here> | Rating: <criteria evaluation result here>. Users want to know more about how this rating is justified in the privacy policy. When answering the questions, focus on the given topic of the rating. Keep it short! <Complexity and answer length according to settings>
\\
\textbf{System prompt general chat: }You are an expert in data protection with many years of experience in consumer protection. You have analyzed the following privacy policy and are aware of its risks and ethical implications for users: <Privacy policy here>. 
You should advise users and explain the implications for them in a conversation. <Complexity and answer length according to settings>

\end{document}